# Graph-Structured Deep Learning Framework for Multi-task Contention Identification with High-dimensional Metrics

Xiao Yang
Santa Clara University
Santa Clara, USA

Yinan Ni
University of Illinois at Urbana-Champaign
Urbana, USA

Yuqi Tang
New York University
New York, USA

Zhimin Qiu
University of Southern California
Los Angeles, USA

Chen Wang
University of Missouri–Kansas City
Kansas City, USA

Tingzhou Yuan*
Boston University
Boston, USA

*Abstract*-This study addresses the challenge of accurately identifying multi-task contention types in high-dimensional system environments and proposes a unified contention classification framework that integrates representation transformation, structural modeling, and a task decoupling mechanism. The method first constructs system state representations from high-dimensional metric sequences, applies nonlinear transformations to extract cross-dimensional dynamic features, and integrates multiple source information such as resource utilization, scheduling behavior, and task load variations within a shared representation space. It then introduces a graph-based modeling mechanism to capture latent dependencies among metrics, allowing the model to learn competitive propagation patterns and structural interference across resource links. On this basis, task-specific mapping structures are designed to model the differences among contention types and enhance the classifier's ability to distinguish multiple contention patterns. To achieve stable performance, the method employs an adaptive multi-task loss weighting strategy that balances shared feature learning with task-specific feature extraction and generates final contention predictions through a standardized inference process. Experiments conducted on a public system trace dataset demonstrate advantages in accuracy, recall, precision, and F1, and sensitivity analyses on batch size, training sample scale, and metric dimensionality further confirm the model's stability and applicability. The study shows that structured representations and multi-task classification based on high-dimensional metrics can significantly improve contention pattern recognition and offer a reliable technical approach for performance management in complex computing environments.

## I. Introduction

In contemporary computing environments, system scales continue to expand, the number of components grows rapidly, and task types become increasingly diverse[1]. These factors make resource contention more concealed and dynamic in high-dimensional settings. With the widespread adoption of cloud computing, distributed scheduling, high-concurrency services, and multi-tenant architectures, performance bottlenecks are no longer caused by shortages of a single resource. Instead, they arise from the amplified interactions among multi-source metrics, the heterogeneity of task behaviors, and the accumulation of environmental uncertainty. In high-dimensional metric spaces, traditional approaches based on single-dimensional or low-dimensional correlations cannot accurately describe the formation and impact of contention. System states often exhibit nonlinear, dependent, and multi-scale coupling properties, which makes contention identification and classification a core challenge in modern systems management.

As monitoring granularity becomes more fine-grained, metrics now span CPU, memory, disk, and network, as well as application-level indicators such as queue length, thread saturation, cache hit rate, and lock wait time. They also include platform-level signals such as scheduling delay and load balancing states. Together, these form a high-dimensional and cross-layer metric spectrum. The metrics show strong temporal dependencies and often involve local disturbance amplification, chained blocking propagation, and repeated competitive patterns. These characteristics form typical multi-task contention scenarios. Under such conditions, effective modeling of high-dimensional metrics, identifying hidden contention patterns in multi-source data, and distinguishing different contention types become urgent problems in both theory and engineering. The rise in task concurrency and service complexity further increases the difficulty of contention classification. The challenge comes not only from data dimensionality but also from the diversity of scenarios and the uncertainty of system behaviors[2].

Meanwhile, application systems are rapidly moving toward intelligent and automated operational paradigms. Predictive scheduling, proactive autoscaling, assisted root-cause localization, and adaptive policy tuning are evolving. Systems must not only observe the consequences of contention but also detect contention types and potential triggers in advance. This capability supports real-time or near-real-time policy responses. Under this trend, contention classification based on high-dimensional metrics is no longer a simple performance analysis task[3,4]. It has become a foundational module of intelligent operations. It is essential for building platforms capable of self-monitoring, self-diagnosis, and self-optimization. Only when contention types are accurately identified can systems form effective control loops in resource management, scheduling, throttling, and priority adjustment[5].

Contention classification based on high-dimensional metrics also carries deeper research significance. It can reveal the structural characteristics behind multi-task competition and reflect system pressure patterns, bottleneck structures, and cross-resource coupling[6]. A systematic analysis of these relationships advances theoretical understanding of competitive behaviors in high-dimensional systems. It enables researchers to characterize system stability, resilience, and sensitivity under heavy load from multiple perspectives. Accurate distinction among contention types also provides quantitative support for automated operations, capacity planning, fault prediction, and strategy optimization. It promotes a shift from passive response to proactive governance[7].

## II. RELATED WORK

Recent research in high-dimensional system management has seen the emergence of a variety of deep learning approaches and advanced representation learning techniques to address the complexity of modern computing environments. A foundational line of work has shown that capturing both temporal dependencies and multi-scale structures is essential for robust anomaly detection in high-dimensional metrics. Kang [7] proposed a deep learning-based framework that integrates temporal and structural awareness for anomaly detection, providing crucial insights into handling rich metric data. Extending this perspective, Cheng [9] introduced hierarchical attention mechanisms to model latent dependencies among metrics, further demonstrating the benefit of structured, multi-level modeling for dynamic system states.

Approaches based on reinforcement learning and adaptive scheduling have also become important in dynamic, resource-intensive environments. Gao et al. [10] utilized deep Q-learning to address scheduling challenges, emphasizing adaptive policy learning in complex system environments. Zhou [11] contributed a unified reinforcement learning architecture focused on adaptive representation transfer and multi-task policy optimization—core strategies that underpin many modern classification frameworks. Techniques for representation learning and transfer have also been widely adopted to improve generalization across diverse tasks. Zhou [12] explored self-supervised transfer learning with shared encoders, which provides methodological support for unified frameworks capable of extracting shared and task-specific features from high-dimensional data. Such transfer mechanisms are especially relevant in multi-source, heterogeneous scenarios like system contention classification.

In the broader deep learning landscape, meta-learning and efficient fine-tuning methods have advanced the robustness and adaptability of models facing evolving patterns or sample scarcity. Hanrui et al. [13] introduced meta-learning techniques for adaptive feature extraction and rapid task adjustment, reflecting principles that can be extended to multi-task contention identification. Parameter-efficient fine-tuning [14] and confidence-driven multi-granular retrieval [15] both contribute strategies for scalable, robust feature learning and reliable prediction under uncertain, high-dimensional conditions. Techniques such as risk quantification and uncertainty modeling [16] have also enhanced model reliability, which is critical for performance management systems that operate in noisy and volatile environments. Additionally, the use of graph-based modeling and multi-scale adaptation continues to shape methodologies for high-dimensional data integration and dependency discovery. Wu et al. [17] demonstrated graph integration in their transformer architecture, and Zhang et al. [18] explored multi-scale adaptation strategies for efficient representation learning. These works underscore the value of structural modeling and scalable adaptation in handling complex system interactions.

Together, these studies form the methodological backbone for the present work, which integrates structured representation transformation, graph-based dependency modeling, and adaptive multi-task learning to address the challenges of high-dimensional contention classification. By drawing from deep metric modeling, reinforcement learning, transfer learning, meta-learning, and scalable adaptation, this research extends the current state-of-the-art in system performance analysis and resource contention identification.

## III. PROPOSED FRAMEWORK

The methodological framework of this study aims to extract stable contention structure features from a high-dimensional system indicator space and achieve high-precision classification of different contention types based on a multi-task modeling mechanism. The overall architecture of the model is shown in Figure 1.

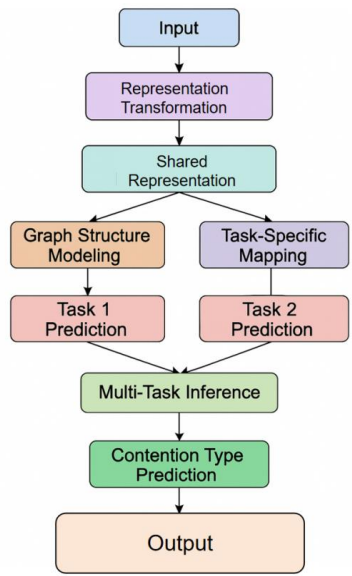

Figure 1. Overall architecture diagram

First, the original monitoring sequence is represented as a high-dimensional tensor containing multi-source indicators, denoted as:

$$X \in R^{T \times D} \quad (1)$$

Where T represents the time duration and D represents the metric dimension. To capture the inherent correlations between different metrics, a representation transformation network is introduced to map the input to a contention-based representation space:

$$H = f_\theta(X) \quad (2)$$

Where $f_\theta$ is a trainable nonlinear transformation function. The design goal of this representation space is to preserve the dynamic characteristics of the system while reducing irrelevant disturbances through noise reduction and structural purification mechanisms, so that the representation H can explicitly characterize potential resource competition paths and patterns.

After deriving the shared representation, the system further models the multi-task structure of competitive behavior using multiple parallel prediction heads. Informed by the approach of Zhang et al.[19], an unsupervised framework is employed in which each task is handled by a dedicated prediction head, effectively capturing temporal dependencies and service-specific variations in multi-source cloud environments. To address the risk of gradient conflict and model bias across different tasks, Gao et al.'s parameter-decoupled multi-task architecture is adopted [20]. This method assigns the shared representation to task-specific mapping functions, ensuring independent parameter updates and mitigating interference between tasks during learning. Furthermore, the adoption of deep temporal convolutional neural networks with attention mechanisms, as demonstrated by Lyu et al. [21], enables the architecture to dynamically focus on relevant features within the high-dimensional metric sequences, further enhancing contention classification accuracy across diverse resource scenarios. The parameter-decoupled multi-task mapping is thus formalized as:

$$z_k = g_k(H), \quad k = 1,2,\ldots K \quad (3)$$

Each $g_k$ corresponds to a discrimination structure for a specific contention type. This construction enables the classifier to learn differentiated features for scenarios such as CPU contention, I/O conflicts, memory pressure, network bottlenecks, and cross-resource hybrid contention. To maintain a learning balance across different tasks during optimization, an adaptive weighting mechanism is introduced, and the overall objective is constructed as follows:

$$L_{multi} = \sum_{k=1}^{K} w_k L_k \quad (4)$$

Where $w_k$ is the dynamically adjusted task weight, and $L_k$ is the loss function for task k.

In the representation learning phase, to enhance the model's ability to perceive the structure of high-dimensional resource indicators, an implicit association modeling mechanism based on graph structure is introduced. The coupling relationships between high-dimensional indicators are represented as a graph:

$$G = (V, E) \quad (5)$$

Here, the vertex set V corresponds to the index dimension, and the edge set E represents the potential competition or dependency relationships between resources. After graph construction, a structure-enhanced representation is obtained through graph structure transformation:

$$\widetilde{H} = \phi(G, H) \quad (6)$$

Here, $\phi(\cdot)$ represents a graph-based propagation-based structure mapping operation. This structure enables the model to capture cascading effects and chained contention across resources, thereby improving its ability to identify complex contention patterns.

In the final inference phase, multi-task prediction results are integrated into a unified contention discrimination framework. To ensure the comparability of different task outputs in the decision space, a normalized contention confidence score is introduced, defined as:

$$p_k = \frac{exp(z_k)}{\sum_{j=1}^{K} exp(z_j)} \quad (7)$$

Here, $p_k$ reflects the probability that the system belongs to the k-th type of contention mode at the current moment. Based on the above structure, the model forms a complete method chain consisting of high-dimensional representation transformation, structural modeling, multi-task discrimination, dynamic weight optimization, and unified inference, enabling it to identify fine-grained contention types in complex system environments while maintaining robustness to high-dimensional index noise and dynamic perturbations.

IV. EXPERIMENTAL ANALYSIS

*A. Dataset*

This study uses the publicly available Alibaba Cluster Trace 2018 dataset as the source of system-level multi-metric inputs. The dataset contains real operational records collected from a large-scale distributed cluster. It includes mixed executions of a large number of online and offline tasks and provides rich monitoring metrics at the machine, container, and task levels. The dataset records time series traces of CPU utilization, memory usage, disk read and write rates, network traffic, scheduling wait times, task lifecycle information, and the differences between requested and actual resource usage. These metrics allow a comprehensive description of resource contention patterns in modern computing systems.

Each machine in the dataset contains multi-dimensional performance metrics. The recording frequency is high, and the data span is long. These characteristics capture the complex and dynamic structure of resource contention in real production environments. The relationships among task types, the changes in load, and the behaviors of resource over-allocation and under-allocation form typical multi-task contention scenarios. Because the metrics are high-dimensional, long-term, and strongly cross-correlated, the dataset is well-suited for high-dimensional metric modeling, complex contention pattern mining, and multi-task classification research.

The openness and reproducibility of Alibaba Cluster Trace 2018 provide a unified comparative environment for system performance studies. Models can be validated on representative real platform data. Its high-dimensional, multi-source, heterogeneous, and strongly dynamic properties make the dataset widely useful for research on contention type identification, scheduling behavior modeling, and system bottleneck analysis. Therefore, it offers a reliable data foundation for the design and evaluation of multi-task contention classification algorithms.

*B. Experimental Results*

We first benchmark the proposed model against several representative baselines, and the comparative results are summarized in Table 1.

Table 1. Comparative experimental results

| Method | Acc | Recall | Precision | F1-Score |
|---|---|---|---|---|
| MLP[22] | 0.842 | 0.817 | 0.823 | 0.820 |
| XGBoost[23] | 0.874 | 0.861 | 0.846 | 0.853 |
| GNN[24] | 0.892 | 0.879 | 0.867 | 0.873 |
| GAT[25] | 0.903 | 0.888 | 0.881 | 0.884 |
| Ours | 0.932 | 0.918 | 0.907 | 0.912 |

The comparative results show that performance consistently improves with increasing model complexity and representation capacity, indicating that contention patterns in high-dimensional system metrics exhibit strong structural properties and cross-dimensional dependencies. Traditional MLP models perform poorly due to their reliance on fixed transformations and inability to capture nonlinear coupling among metrics, while XGBoost benefits from feature splitting but remains limited in modeling dynamic and graph-structured dependencies. GNNs significantly improve performance by explicitly modeling cross-metric propagation and structural interactions, and GAT further enhances precision and recall through attention-based weighting of critical dependencies. The proposed multi-task graph method achieves the best results across all metrics, including F1, by combining shared structural representations with task-specific prediction heads to capture diverse contention behaviors, demonstrating superior feature extraction and robustness in complex, dynamic system environments; the impact of batch size on performance is further analyzed in Figure 2.

The results show that batch size has a clear nonlinear effect on contention classification performance in high-dimensional system metrics, reflecting a trade-off between gradient stability and sensitivity to structural details. Small batch sizes (e.g., 16) capture fine-grained fluctuations but suffer from high gradient variance, leading to slightly lower accuracy and F1, while moderate batch sizes (32–64) yield more stable gradients and more coherent shared representations, producing consistent improvements across all metrics, especially accuracy and recall. When the batch size increases to 128, performance degrades, indicating that overly large batches oversmooth gradients, reduce sensitivity to local structural changes and transient contention bursts, and weaken the model's ability to distinguish subtle inter-task differences. Overall, accuracy, recall, precision, and F1 all indicate that moderate batch sizes offer the best balance between stability and representational capacity, highlighting the importance of careful batch-size selection for multi-task, high-dimensional contention modeling; the sensitivity to training-set size is further analyzed in Figure 3.

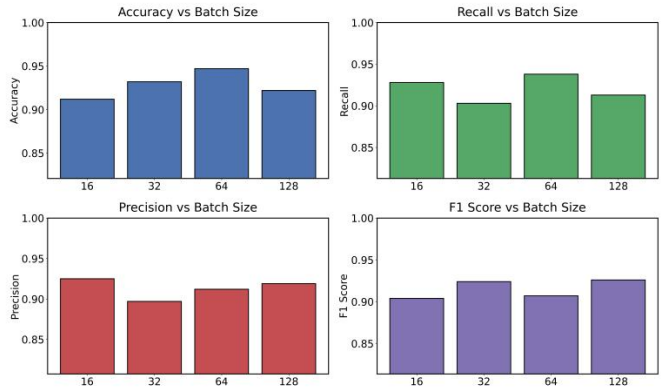

Figure 2. The effect of batch size on experimental results

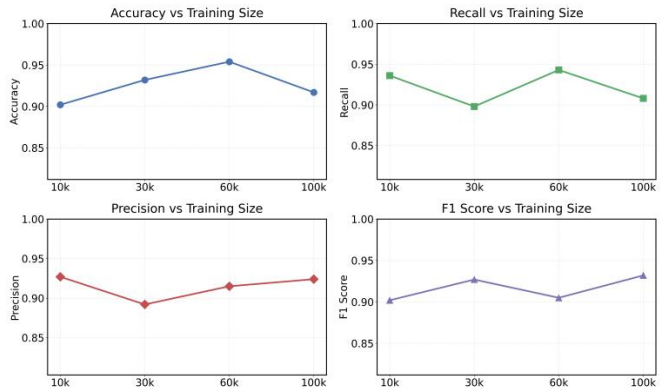

Figure 3. The impact of training set sample size on experimental results

Training-set size has a clear non-monotonic effect on model performance. With a small dataset, the model captures only coarse patterns and lacks coverage of complex contention behaviors, resulting in moderate accuracy and F1. Performance improves as data scale increases, peaking at a moderate size where diverse contention patterns are well represented. However, excessively large datasets slightly degrade performance by introducing low-contribution samples and oversmoothing fine-grained contention features, indicating that an appropriate balance in data scale is critical for optimal classification.

V. CONCLUSION

This study focuses on the problem of multi-task contention classification under high-dimensional system metrics and builds a unified framework that models both shared structures and task-specific differences. By incorporating high-dimensional representation transformation, graph-based modeling, and task decoupled classification strategies, the model maintains stable recognition performance even when metric correlations are complex, resource competition patterns are diverse, and system dynamics are highly unstable. The results show that contention patterns exhibit strong cross-dimensional dependencies and structural coupling. A single type of representation or a single task model cannot fully capture the complex competitive behaviors within the system. The proposed framework strengthens the theoretical understanding of contention mechanisms in high-dimensional systems and advances system behavior modeling from a multi-task perspective.

At the application level, this study provides a transferable technical basis for cloud scheduling, distributed resource management, service quality assurance, and performance governance in multi-tenant systems. With more detailed and structured contention classification, systems can identify bottleneck sources more precisely and support smarter automated scheduling and optimization strategies. The method is also valuable for capacity planning, anomaly detection, root cause analysis, and system health prediction. It offers new analytical tools for improving the stability and controllability of large-scale systems. As system scale continues to grow and resource competition intensifies, algorithms capable of distinguishing contention patterns at fine granularity will become essential components for ensuring stable operation in critical services. Future work can explore the model's generalization ability in highly dynamic environments, hybrid cloud settings, and higher-dimensional metric systems. It can also integrate online learning, reinforcement-based scheduling, and adaptive optimization mechanisms so that contention identification can operate within real-time system control loops. In addition, the structural awareness of the model can be extended to learn complex topologies, cross-node dependencies, and chained bottleneck propagation paths. These advances can support higher autonomy and evolution in intelligent system management. Continued improvements in modeling high-dimensional system metrics have the potential to play a central role in future automated operations, intelligent scheduling engines, and large-scale resource management platforms.